\documentclass[aps,pra,twocolumn,showpacs,letterpaper,superscriptaddress,longbibliography]{revtex4-1}

\usepackage[colorlinks=true, citecolor=blue, linkcolor=blue, urlcolor=blue]{hyperref}
\usepackage{graphicx,dcolumn,longtable,epsfig}
\usepackage[usenames]{color}
\usepackage{amssymb}
\usepackage{amsmath}
\usepackage{bm}
\usepackage{footnote}
\usepackage{float}
\usepackage{subfigure}
\usepackage{color}

\usepackage[english]{babel}

\usepackage{ulem}
\usepackage[T1]{fontenc}

\newcommand{\be}{\begin{equation}}
	\newcommand{\ee}{\end{equation}}

\def\bea{\begin{eqnarray}}
	\def\eea{\end{eqnarray}}

\begin{document}
	
	\title{Bose-Hubbard Model on a Honeycomb Superlattice: Quantum Phase Transitions and Lattice Effects}
	
	\author{Wei-Wei Wang}
	\affiliation{Department of Physics, Anhui Normal University, Wuhu, Anhui 241000, China}
	\author{Jin Yang}
	\email{jinyang@ahnu.edu.cn}
	\affiliation{Department of Physics, Anhui Normal University, Wuhu, Anhui 241000, China}
	\author{Jian-Ping Lv}
	\email{jplv2014@ahnu.edu.cn}
	\affiliation{Department of Physics, Anhui Normal University, Wuhu, Anhui 241000, China}
	\affiliation{Department of Physics, Anhui Province Key Laboratory for Control and Applications of Optoelectronic Information Materials, Key Laboratory of Functional Molecular Solids, Ministry of Education, Anhui Normal University, Wuhu, Anhui 241000, China}
	\author{Chao Zhang}
	\email{chaozhang@ahnu.edu.cn}
	\affiliation{Department of Physics, Anhui Normal University, Wuhu, Anhui 241000, China}
	
	\begin{abstract}
		We investigate the ground-state and finite-temperature phase diagrams of the Bose-Hubbard model on a honeycomb superlattice. 
		The interplay between the superlattice potential depth $\Delta/t$ and the onsite interaction $U/t$ gives rise to three distinct quantum phases at zero temperature: a superfluid phase, a Mott insulator I phase with unit filling on each site, and a Mott insulator II phase characterized by density imbalance—double occupancy on one sublattice and vacancy on the other at unit filling. The SF–MI transitions are found to be continuous, consistent with second-order quantum phase transitions. We further extend our analysis to finite temperatures within the superfluid regime. Our work highlights how a honeycomb superlattice geometry enables access to interaction- and lattice-modulation-driven quantum phases, including a density-imbalanced Mott insulator and a robust superfluid regime, offering concrete theoretical predictions for cold-atom experiments.
	\end{abstract}
	
	\maketitle
	
	\section{INTRODUCTION}
	Honeycomb lattices are bipartite lattices which can be decomposed as two triangular lattices with geometric frustration. This is in profound difference with square lattices, which can be successively seen as combinations of larger square lattices. It is well known that due to electronic structure of carbon atoms, many carbon-based compounds are of honeycomb lattice structures, including intriguing materials like carbon nanotubes and graphene~\cite{iijima1991helical}. The topological properties of graphene and its remarkable band structure lead to many novel quantum phases and the linear dispersion relation at the Dirac points gives rise to phenomena like quasi-relativistic particles and an anomalous quantum Hall effect~\cite{geim2007rise,miao2022bosonic,hirata2021interacting}. Thus, a precise and thorough study of the quantum phase diagram of honeycomb lattices is of significance and necessity. Ultracold atoms in optical lattices offers an excellent platform for exploring quantum phase transitions and many-body physics in lattices~\cite{Jaksch1998,Qian2011,Bloch2008,jaksch2005cold}. As many experiments devote into the study of square lattices, only a few experiments were carried out on honeycomb lattices~\cite{soltan2011multi,soltan2012quantum,Tarruell2012}. These delicately designed experiments smartly utilized the high tunability of optical lattices and explored multi-component effects and multi-orbital effects. Compared to square lattices, the less symmetric honeycomb lattices posed challenges to some extent for both theoretical and experimental studies.

	In this work, we employed the quantum Monte Carlo method with the worm algorithm~\cite{Prokofev1998} to systematically investigate the ground-state and finite-temperature phase diagrams of the Bose-Hubbard model on a honeycomb superlattice. The study on superlattices is being intensified today, and extensive studies on one-dimensional superlattices~\cite{Hen2009, Buonsante2004,Dhar2011,Barmettler2008,Rousseau2006,Grusdt2013,Flesch2008} and two-dimensional square superlattices~\cite{Chen2010,Bibo2020,Hen2010,Hen2009,Chalopin2025} have revealed a rich variety of quantum many-body phenomena. These include the emergence of superlattice-induced Mott insulating phases at integer fillings~\cite{Dhar2011,Chen2010}, reentrant quantum phase transitions~\cite{Danshita2008}, fractional corner charges~\cite{Bibo2020}, and topological edge states~\cite{Grusdt2013}. 
	These exciting results provide strong motivation to extend superlattice physics into more complex lattice geometries,
	where what is of particular interest in this work is the honeycomb superlattice. 
	Unlike idealized bipartite triangular lattices—which are challenging to implement with high fidelity—the superlattice approach offers a practical and robust route to engineering symmetry breaking and staggered potentials in a honeycomb geometry.

	The quantum Monte-Carlo method with the worm algorithm we employed here is a numerical method based on path-integral methods. It is straightforward and convenient when dealing with complex lattice structures and provide efficient and unbiased calculation of large size systems. 
	From our study, we identified three distinct phases: a superfluid (SF) phase, a uniform Mott insulator phase with one particle per site (MI-I), and a sublattice-imbalanced Mott insulator phase (MI-II). We further characterize the nature of the SF–MI transitions and examine the thermal melting of the SF phase. Our results reveal the rich structure of quantum phases enabled by the interplay of interaction and lattice modulation, and provide concrete theoretical guidance for future experiments exploring bosons in honeycomb optical superlattices. The remainder of this paper is organized as follows: In Section~\ref{sec:2} and~\ref{sec:3}, we introduce the model Hamiltonian and order parameters. Section~\ref{sec:4} presents the ground-state phase diagram and quantum phase transitions. Section~\ref{sec:5} discusses finite-temperature phase diagram of the above system. Section~\ref{sec:6} concerns experimental realization. Section~\ref{sec:7} concludes the paper.


	\section{Hamiltonian}
	\label{sec:2}
	
	\begin{figure}[t]
		\includegraphics[width=0.5\textwidth]{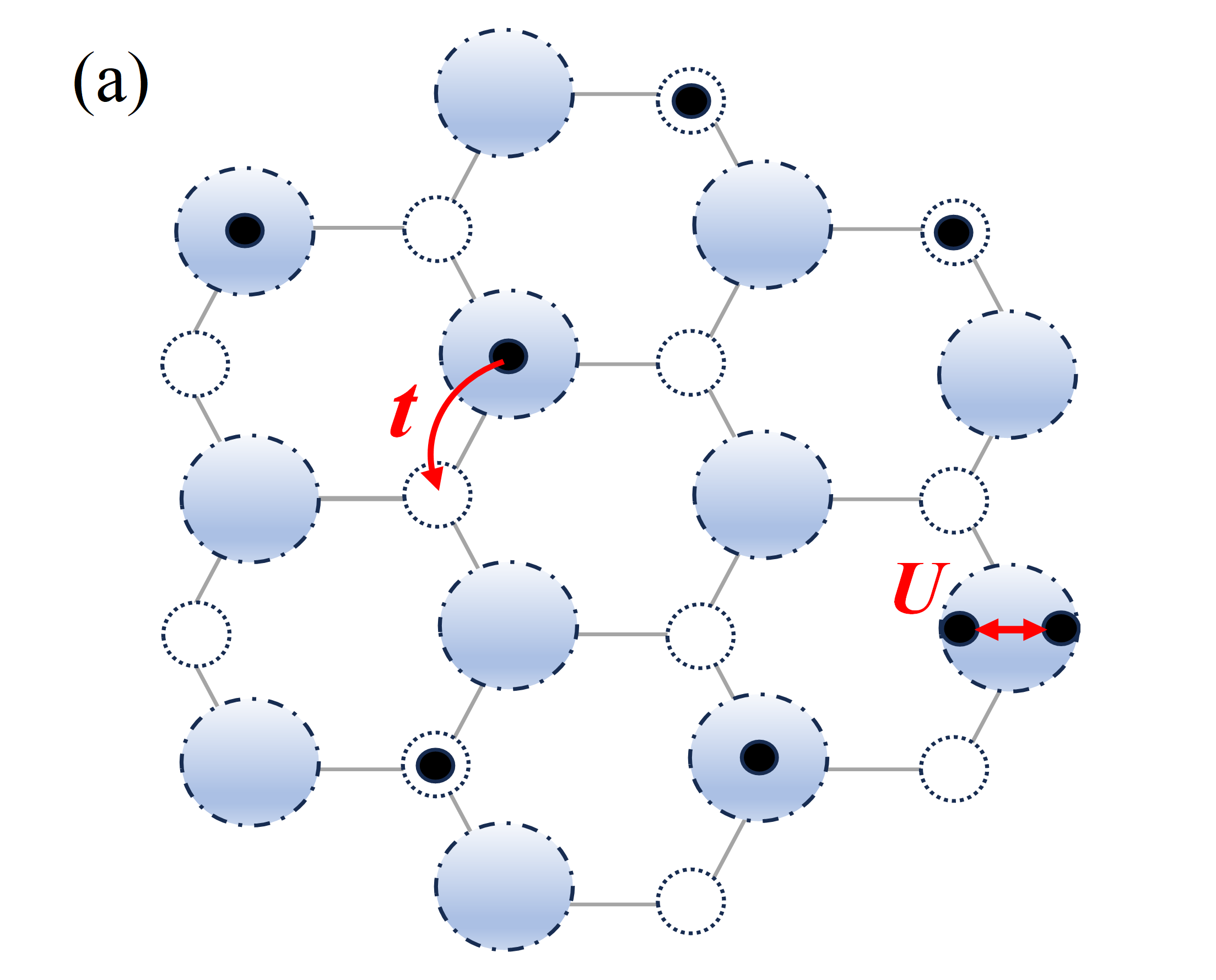}
		\includegraphics[width=0.45\textwidth]{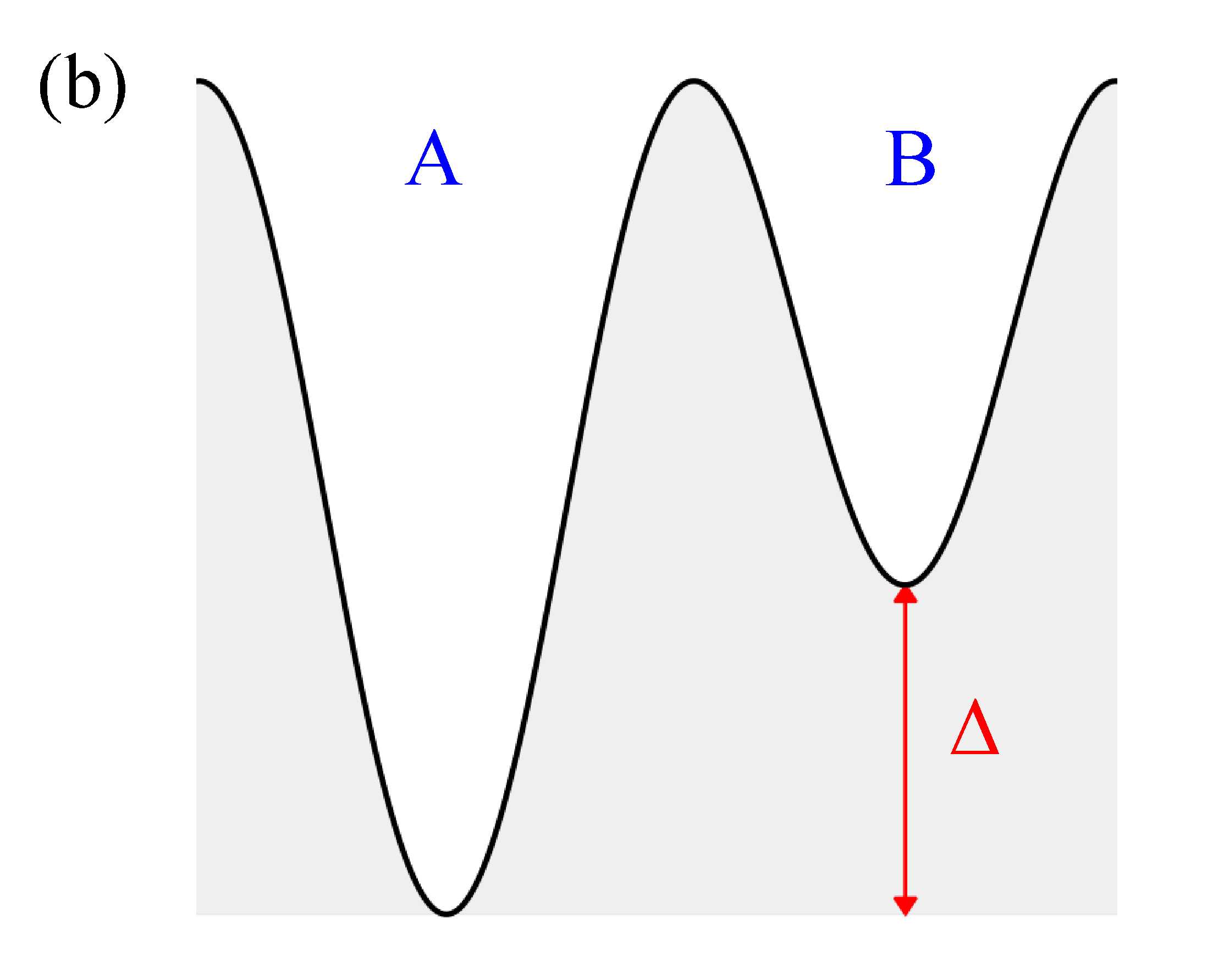}
		\caption{(a) Schematic illustration of the honeycomb superlattice geometry. Blue dashed-dotted circles denote sublattice A, and empty dotted circles denote sublattice B. Bosons can tunnel between nearest-neighbor sites with hopping amplitude $t$. When two bosons occupy the same lattice site, they experience a repulsive onsite interaction $U$. (b) Illustration of the staggered superlattice potential with a bias potential depth $\Delta$, which defines the onsite energy difference between sublattices A and B.}
		\label{setup}
	\end{figure}
	We consider the Bose-Hubbard model defined on a honeycomb superlattice composed of two inequivalent triangular sublattices, labeled A and B, with a staggered onsite potential energy offset $\Delta$. The Hamiltonian of the system is:
	
	\begin{equation}
		\begin{aligned}
			H =\ & -t \sum_{\langle i,j \rangle} \left( b_i^\dagger b_j + \text{h.c.} \right)  + \frac{U}{2} \sum_i n_i(n_i - 1) \\
			& - \Delta \sum_{i \in A} n_i  - \mu \sum_i n_i,
			\label{eq:hamiltonian}
		\end{aligned}
	\end{equation}
	where $b_i^\dagger$ $( b_i )$ is the bosonic creation (annihilation) operator at site $i$, and $n_i = b_i^\dagger b_i$ is the number operator. The first term describes nearest-neighbor hopping with amplitude $t$, the second term is the onsite repulsive interaction with strength $U > 0$, the third term introduces a staggered potential $\Delta$ applied to the sublattice A, and the last term couples to the total particle number via the chemical potential $\mu$. 
	
	The honeycomb lattice is bipartite, with each site connected to three nearest neighbors from the opposite sublattice~\cite{koziol_quantum_2024,lado2024}. The superlattice potential depth $\Delta$ explicitly breaks sublattice symmetry, and together with $U/t$ and $\mu/t$, controls the ground-state properties. At commensurate fillings, the system can exhibit insulating phases stabilized by interaction and lattice modulation. Figure~\ref{setup} illustrates the honeycomb superlattice structure composed of two inequivalent sublattices, labeled A and B, arranged in a hexagonal geometry. A staggered potential is applied such that sites on sublattice A experience an additional onsite energy $\Delta$, while sublattice B sites remain at zero energy offset. This staggered modulation effectively forms a periodic double-well potential within each unit cell and explicitly breaks the sublattice symmetry. Bosons are allowed to tunnel between adjacent sites with hopping amplitude $t$, and interact via an onsite repulsion $U$ when multiple bosons occupy the same site.  
	The energy offset $\Delta$ between sites A and B governs the density imbalance between the two sublattices and plays a key role in stabilizing distinct quantum phases such as the superfluid and two types of Mott insulating states explored in this work.

	\section{Method and order parameters}
	\label{sec:3}
	We investigate this system using path-integral quantum Monte Carlo simulations with the worm algorithm~\cite{Prokofev1998}. The key order parameters we measure are the superfluid density, compressibility, structure factor, and sublattice density imbalance.
	
	\textit{Superfluid Density:}
	The superfluid density $\rho_s$ is calculated in terms of the winding number fluctuations~\cite{Pollock1987}:
	$\rho_s = \frac{\langle \mathbf{W}^2 \rangle}{d L^{d-2} \beta}$,
	where $\langle \mathbf{W}^2 \rangle = \sum_{i=1}^{d} \langle W_i^2 \rangle$ is the expectation value of the winding number square, $d = 2$ is the spatial dimension, $L$ is the linear system size, and $\beta$ is the inverse temperature. A finite $\rho_s$ indicates phase coherence and superfluidity.
	
	\textit{Compressibility:}
	The compressibility $\kappa$ is defined as:
	$\kappa = \frac{\beta}{L^2} \left( \langle N^2 \rangle - \langle N \rangle^2 \right)$,
	where $N$ is the total number of particles. Compressibility is finite in SF phase and vanishes (in the thermodynamic limit) in incompressible Mott insulator phases.
	
	\textit{Structure Factor:}
	The structure factor $S(\mathbf{k})$ characterizes diagonal long-range order (density modulation) and is defined as:
	$S(\mathbf{k}) = \frac{1}{N} \sum_{\mathbf{r}, \mathbf{r}'} e^{i \mathbf{k} \cdot (\mathbf{r} - \mathbf{r}')} \langle n_{\mathbf{r}} n_{\mathbf{r}'} \rangle$,
	where $N$ is the total number of particles and $\mathbf{k}$ is the reciprocal lattice vector. We use $\mathbf{k} = (2\pi/3, 2\pi/3)$ to detect the density wave order. However, since such modulation is intrinsic to the honeycomb lattice geometry, the structure factor alone is insufficient to distinguish the two Mott insulator phases.
	
	\textit{Sublattice Density Imbalance:}
	To differentiate the two Mott insulating phases (MI-I and MI-II), we define the sublattice density imbalance:
	$\Delta n = \left| \langle n_A \rangle - \langle n_B \rangle \right|$,
	where $\langle n_A \rangle$ and $\langle n_B \rangle$ are the average particle densities on sublattices A and B, respectively:
	$\langle n_A \rangle = \frac{1}{N_A} \sum_{i \in A} \langle n_i \rangle, \quad
	\langle n_B \rangle = \frac{1}{N_B} \sum_{i \in B} \langle n_i \rangle$. $N_A$, $N_B$ is the number of sites in sublattice A and B. In the uniform Mott insulator (MI-I), $\Delta n = 0$, while in the sublattice-imbalanced Mott insulator (MI-II), $\Delta n$ is finite due to the presence of density imbalance induced by the superlattice potential.

	In all our simulations, the system sizes used are $L = 12$, 15, 18, 21, 24, and up to $L = 36$. We set the inverse temperature as $\beta = 1/T = L$, with $T$ denoting the temperature. This choice is consistent with a dynamical critical exponent of $z = 1$, ensuring that the simulations effectively capture ground-state behavior near quantum criticality.

	\section{GROUND-STATE PHASE DIAGRAM}
	\label{sec:4}
	
	\begin{figure}[t]
		\centering
		\includegraphics[width=0.45\textwidth]{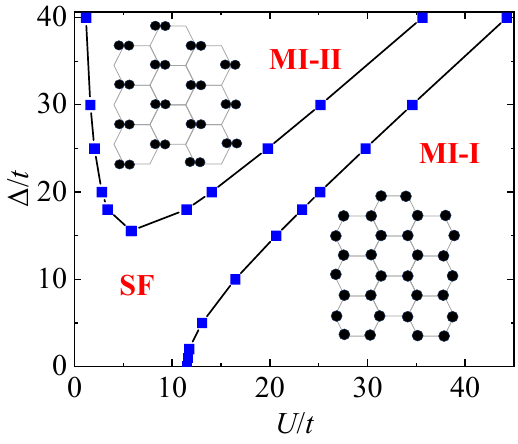}
		\caption{Phase diagram of the system described by Eq.~\ref{eq:hamiltonian} at filling factor $n = 1$. The horizontal and vertical axes represent the onsite interaction strength $U/t$ and the superlattice potential depth $\Delta/t$, respectively. As these two parameters are varied, the system exhibits three distinct phases: a superfluid (SF), a uniform Mott insulator (MI-I) with one particle per site, and a sublattice-imbalanced Mott insulator (MI-II) characterized by double occupation on one sublattice and vacancy on the other. Solid blue squares indicate the numerically determined phase boundary between the SF and MI phases. The solid black line is a guide to the eye. The boundary between MI-I and MI-II is determined scaling the superfluid density and monitoring the sublattice density imbalance. Error bars are smaller than the symbol size when not visible.}
		\label{fig:phase}
	\end{figure}

	In the following, we present a numerical study of the Hamiltonian~\ref{eq:hamiltonian} at unit filling ($n = N / N_{\text{site}} = 1$) using path-integral quantum Monte Carlo simulations based on the worm algorithm. The honeycomb superlattice features a hexagonal geometry, where each site is connected to three nearest neighbors. While the lattice structure is non-orthogonal and differs from square lattices in coordination and symmetry, we simulate the system on a regular grid consisting of $N \times N$ sites. The physical coordinates of the sites follow the honeycomb geometry, but the total number of sites remains the same as in a square lattice, ensuring consistency in system size across different geometries.

	\begin{figure*}[t]
		\centering
		\includegraphics[width=1\textwidth]{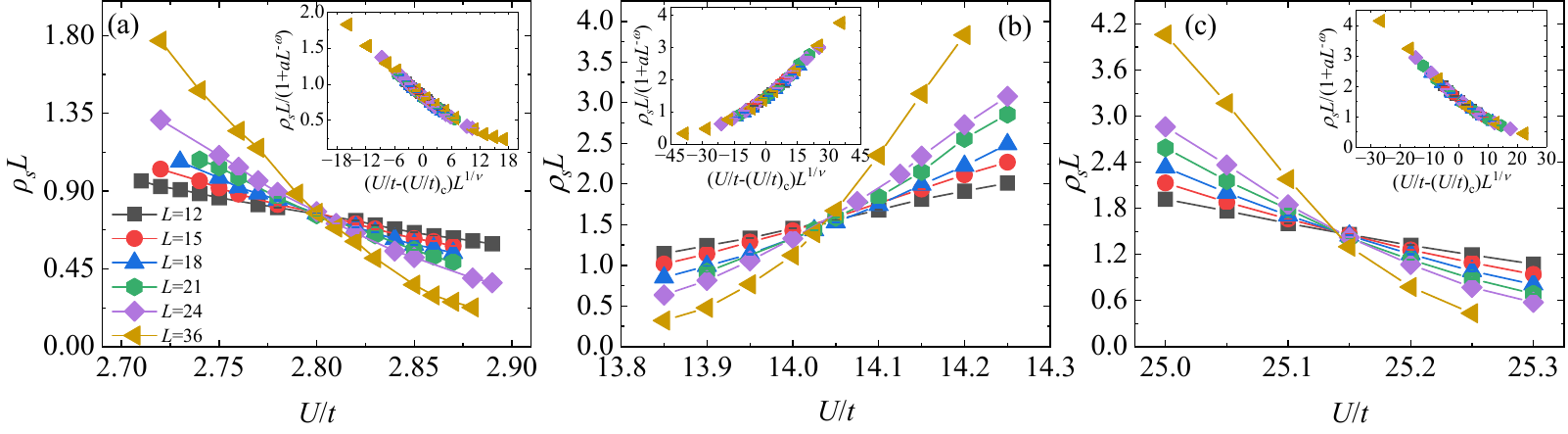}
		\caption{Scaling behavior of the superfluid density $\rho_{s}L$ as a function of the interaction strength $U/t$ for the honeycomb superlattice Bose--Hubbard model at unit filling and fixed superlattice potential $\Delta/t = 20$. Here, $\Delta$ denotes superlattice potential depth. The main panels display the quantum Monte Carlo data for various linear system sizes $L = 12, 15, 18, 21, 24$, and $36$, while the insets illustrate the corresponding data collapse indicating universal scaling near the critical points. (a) SF to MI-II phase transition with critical point $(U/t)_c = 2.7970 \pm 0.0005$. (b) MI-II to SF phase transition with critical point $(U/t)_c = 14.0381 \pm 0.0005$. (c) SF to MI-I phase transition with critical point $(U/t)_c = 25.1357 \pm 0.0005$.}
		\label{datacollapse} 
	\end{figure*}

	The ground-state phase diagram of the honeycomb superlattice at unit filling is shown in Fig.~\ref{fig:phase}, where the horizontal and vertical axes represent the onsite interaction strength $U/t$ and the superlattice potential depth $\Delta/t$, respectively. At $\Delta/t = 0$, the system reduces to a uniform honeycomb lattice. In this limit, the system remains in the superfluid (SF) phase for $U/t \lesssim 11.5$, consistent with previous studies~\cite{Teichmann2010,Luhmann2013}.

	The introduction of a superlattice potential depth $\Delta$ explicitly breaks the translational symmetry of the system by introducing a periodic modulation of the onsite chemical potential. In the honeycomb superlattice, this results in a two-sublattice with a triangular structure (A and B) with alternating onsite energies. Consequently, the effective chemical potential varies periodically, giving rise to insulating phases that are not purely interaction-driven but are stabilized by the lattice modulation. These phases are referred to as superlattice-induced Mott insulators~\cite{Dhar2011}.
	
	We identify two distinct superlattice-induced Mott insulating phases, denoted MI-I and MI-II, depending on the density distribution between the sublattices. In the MI-I phase, the average occupation on both sublattices remains approximately equal to unity ($\langle n_A \rangle \approx \langle n_B \rangle \approx 1$), but particle mobility is suppressed due to the interplay of interactions and weak superlattice modulation (see inset of Fig.~\ref{fig:phase}). In contrast, the MI-II phase emerges at large $\Delta/t$ and $U/t$, characterized by a significant density imbalance between sublattices. Specifically, the deeper sublattice becomes doubly occupied ($\langle n_A \rangle \approx 2$), while the shallower sublattice becomes nearly empty ($\langle n_B \rangle \approx 0$). To quantitatively distinguish between these two phases, we monitor the sublattice density imbalance $\Delta n = |\langle n_A \rangle - \langle n_B \rangle|$ which remains zero in MI-I but becomes finite in MI-II. Although both MI-I and MI-II phases are incompressible and gapped, their physical origins differ significantly. MI-I resembles a conventional Mott insulator primarily stabilized by strong onsite interactions, while MI-II results from a competition between strong interactions and a large staggered potential.
	
	From the phase diagram, distinct phase behavior emerges depending on the strength of the superlattice potential depth $\Delta/t$. At small values of $\Delta/t$, the system exhibits a second-order quantum phase transition from the SF to the MI-I phase. The continuous nature of this transition is confirmed by a detailed finite-size scaling analysis of the superfluid density $\rho_s$, which smoothly vanishes at the critical boundary. At larger values of $\Delta/t$, the strong superlattice potential depth induces the transition into the MI-II phase with pronounced sublattice density imbalance. Between these two insulating phases, an intermediate regime exists in which the competition between the onsite interaction $U/t$ and the superlattice potential depth $\Delta/t$ stabilizes a robust SF phase. In this intermediate region, bosons coherently delocalize across the lattice, resisting localization driven by either strong onsite interactions or superlattice potential depth.
	
	\begin{table*}[h]
		\begin{center}
			\renewcommand{\arraystretch}{1.5} 
			\caption{Fits of $\Delta/t = 20$ data to Eq.~\ref{eq:4} with the correction exponent $\omega = 0.789$ is adopted. In the critical SF to MI-II phase transition, Eq.~\ref{eq:4} was fitted up to the $a_3$ term, yielding $a_3 = 9.71 \times 10^{-6}$ with an error of $1.22 \times 10^{-6}$. For the reverse MI-II to SF critical transition, fitting to the $a_4$ term gave $a_3 = -4.05 \times 10^{-6}$ (error: $0.12 \times 10^{-6}$) and $a_4 = -5.56 \times 10^{-8}$ (error: $0.39 \times 10^{-8}$). In the SF to MI-I critical transition, fitting up to the $a_3$ term yielded $a_3 = 3.93 \times 10^{-6}$ with an error of $0.48 \times 10^{-6}$. }
			\label{fit_result}
			\begin{tabular}[t]{p{2.5cm}p{1.8cm}p{1.8cm}p{1.8cm}p{1.8cm}p{1.5cm}p{1.8cm}p{1.8cm}p{1.8cm}}
				\hline
				\hline
				Phase Transition   &$(U/t)_c$  & $\chi^2$/DF  & $v$ & $a$ & $\omega$ & $Q_0$ & $a_1$ & $a_2$   \\
				\hline
				SF---MI-II       & 2.7970(5)  & 36.5/65     & 0.674(2)  & -0.505(31)   & 0.789   & 0.825(5)      & -0.0547(8)     & 0.001006(3)     \\
				MI-II---SF       & 14.0318(5) & 15.2/42     & 0.669(2)  & 0.364(26)    & 0.789     & 1.448(6)   & 0.0521(6)      &0.000528(10)                                       \\
				SF---MI-I      & 25.1357(5) & 11.1/34    & 0.679(2)  & -0.340(28)  & 0.789  & 1.579(8)   & -0.0781(8)      &0.001154(29)       \\
				
				\hline
				\hline
			\end{tabular}
		\end{center}
	\end{table*}		
	

	Figure~\ref{datacollapse} illustrates the finite-size scaling analysis performed to accurately determine the SF–MI phase boundary (solid blue squares in Fig.~\ref{fig:phase}). Here, we plot the scaled $\rho_s L^{(d+z-2)}$ (with dynamical critical exponent $z = 1$) as a function of interaction strength $U/t$ for fixed superlattice potential depth $\Delta/t = 20$. The system sizes used are $L = 12, 15, 18, 21, 24$, and $36$, represented by black squares, red circles, blue triangles, green hexagons, purple diamonds, and yellow triangles, respectively.
	
	A noticeable drift in the intersection points of curves for different system sizes indicates that corrections to the standard finite-size scaling relation
	\begin{equation}
		\begin{aligned}
			\rho_s L^{(d+z-2)} = f\left(L^{1/\nu}(U/t - (U/t)_c),\, \beta L^{-z}\right),
			\label{eq:2}
		\end{aligned}
	\end{equation}
	where $f(x,\text{const})$ is a universal scaling function, must be included to achieve an accurate data collapse. After incorporating these corrections, the modified scaling relation is given by~\cite{Zhang2015}
	\begin{equation}
		\begin{aligned}
			\rho_s L^{(d+z-2)} = (1 + a L^{-\omega})\,f\left(L^{1/\nu}(U/t - (U/t)_c),\, \beta L^{-z}\right).
			\label{eq:3}
		\end{aligned}
	\end{equation}
	The inset of Fig.~\ref{datacollapse} displays the rescaled quantity $L\rho_s / (1 + a L^{-\omega})$ plotted against the scaled interaction parameter $(U/t - (U/t)_c)L^{1/\nu}$. To quantitatively verify the scaling form and extract the critical exponents, we conduct least-squares fits to the data using the fitting function:
	\begin{equation}
		\begin{aligned}
			\rho_s L = (1 + aL^{-\omega})\left[\,Q_0 + \sum\limits_{n=1}^{N} a_n \left((U/t - (U/t)_c)L^{1/\nu}\right)^n\right].
			\label{eq:4}
		\end{aligned}
	\end{equation}
	The parameters obtained from the prefered fits are summarized in Table~\ref{fit_result}. From this procedure, we determine the critical exponent $\nu = 0.674 \pm 0.002$ [Fig.~\ref{datacollapse}(a)], in good agreement with the expected value $\nu \approx 0.672$ for the superfluid-insulator transition of a clean two-dimensional Bose system~\cite{PhysRevB.100.064525}.

	\begin{figure}[b]
		\centering
		\includegraphics[width=0.45\textwidth]{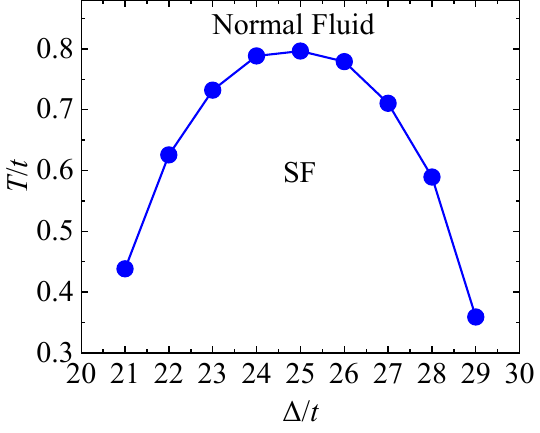}
		\caption{Finite-temperature phase diagram at fixed interaction strength $U/t = 25$, showing the stability region of the SF phase as a function of the superlattice potential depth $\Delta/t$ and temperature $T/t$. The SF phase exists within the range $21 \lesssim \Delta/t \lesssim 29$. The horizontal and vertical axes represent the superlattice potential depth $\Delta/t$ and the scaled temperature $T/t$, respectively. The critical temperature $T_c/t$ marks the boundary between the superfluid and the normal fluid phases, as determined from finite-size scaling analyses.           Error bars are smaller than the symbol size when not visible.
		}\label{T}	 
	\end{figure}

	\section{FINITE-TEMPERATURE RESULTS}
	\label{sec:5}

	In this section, we investigated the robustness of the SF state against thermal fluctuations. Upon increasing the temperature, thermal fluctuations destroy superfluidity via a Berezinskii-Kosterlitz--Thouless (BKT) transition~\cite{Kosterlitz1973}. To systematically analyze this thermal melting, we fix the interaction strength at $U/t=25$ and vary the superlattice potential depth $\Delta/t$ to examine its influence on the critical temperature $T_c$. The dependence of the critical temperature $T_c/t$ on the superlattice potential depth $\Delta/t$ is shown in Fig.~\ref{T}.

\begin{figure}[t]
		\centering
		\includegraphics[width=0.45\textwidth]{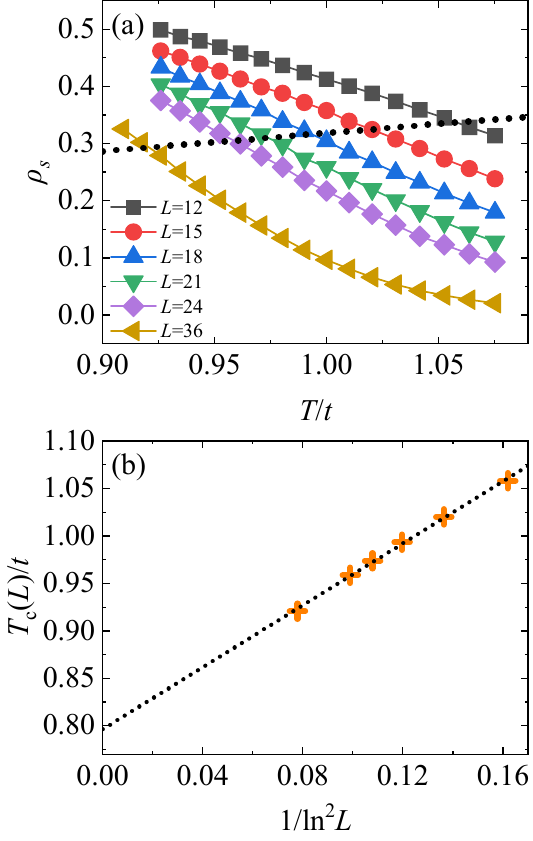}
		\caption{All plots refer to $U/t = 25$ and $\Delta/t=25$. (a) $\rho_s$ as a function of temperature $T/t$ for system sizes $L = 12, 15, 18, 21, 24$, and $36$ (represented by black squares, red circles, blue triangles, green triangles, purple diamonds, and yellow triangles, respectively). The dotted line corresponds to the universal BKT criterion $T/\pi$. The intersection points between this line and the $\rho_s(T)$ curves yield the critical temperatures $T_c(L)/t$ for each finite system size $L$. (b) The critical temperatures $T_c(L)/t$ plotted as a function of $1/\ln^2L$, illustrating the finite-size scaling behavior. Error bars are smaller than the symbol size when not visible.}
		\label{fig_T}
	\end{figure}

	To precisely determine the critical temperature $T_c$, we examine the temperature dependence of $\rho_s$ at fixed particle density $n=1$, interaction strength $U/t=25$, and superlattice potential depth $\Delta/t=25$ for various system sizes $L = 12, 15, 18, 21, 24$, and $36$. In the thermodynamic limit, the BKT transition occurs at a universal jump condition given by $\rho_s(T_c) = \frac{2 m k_B T_c}{\pi \hbar^2}$, where a universal jump in the superfluid density occurs precisely at $T_c$~\cite{Nelson1977,Hebib2024,Zhang2018}. In finite-size simulations, this universal jump is smeared out and becomes rounded, as clearly seen in Fig.~\ref{fig_T}(a).
	
	To systematically extract $T_c$, we adopt the finite-size intersection approach illustrated in Fig.~\ref{fig_T}(a). The dashed line represents the universal BKT criterion, $\rho_s = T/\pi$. We identify the intersection points $T_c(L)/t$ between this line and the finite-size curves of $\rho_s(T)$ for each system size $L$. Extrapolating these intersection points to the thermodynamic limit yields the accurate estimate of the critical temperature $T_c/t$, as presented in Fig.~\ref{fig_T}(b).

	\section{Experimental Realization}
	\label{sec:6}
	
	The Bose-Hubbard model on a honeycomb superlattice can be realized with ultracold bosonic atoms in optical lattices using well-established experimental techniques. Honeycomb lattice geometries have already been successfully engineered in optical lattice experiments through the interference of three coplanar laser beams~\cite{Tarruell2012,Anisimovas2014,Block2010,soltan2011multi,soltan2012quantum}.  
	The onsite interaction $U$ is controlled via Feshbach resonances or by adjusting the lattice depth, while the tunneling amplitude $t$ is determined by the potential barrier between adjacent sites. The sublattice energy difference $\Delta$ can be tuned dynamically by varying the relative phase and amplitude of the superimposed lattices. This enables exploration of the full phase diagram in the $(U/t, \Delta/t)$ parameter space.
	
	The different quantum phases—SF, MI-I, and MI-II—can be detected using a combination of time-of-flight (TOF) imaging, quantum gas microscopy, and Bragg scattering~\cite{hart2015observation}. A foreseen challenge is, unlike the case of square lattices or triangular lattices, sufficiently high potential barriers between adjacent sites of the honeycomb lattices are hard to reach for fixed output power of laser systems in labs. This may preclude the use of the cutting-edge high resolution imaging technics, like quantum gas microscope, which is convenient for detecting MI-I phase and MI-II phase~\cite{asteria2021quantum}. The SF phase can be identified by sharp interference peaks in TOF absorption images, indicating long-range phase coherence. In contrast, the MI-I phase shows a unit filling of one particle per site $\langle n_A \rangle \approx \langle n_B \rangle \approx 1$, with no phase coherence and a flat momentum distribution in TOF.
	The MI-II phase, driven by a strong superlattice potential $\Delta$, exhibits a staggered density pattern with $\langle n_A \rangle \approx 2$ and $\langle n_B \rangle \approx 0$, which can be detected by measuring the sublattice density contrast $\Delta n = |\langle n_A \rangle - \langle n_B \rangle|$ using Bragg scattering. The finite-temperature BKT transition from the SF phase can be observed by monitoring the disappearance of TOF interference peaks and broadening of the momentum distribution. And the realized temperatures can be adjusted by tweaking the evaporation processes.

	\section{Conclusion}
	\label{sec:7}
	In summary, we have systematically investigated the ground-state and finite-temperature phase diagrams of the Bose-Hubbard model on a honeycomb superlattice using path-integral quantum Monte Carlo simulations with the worm algorithm. By tuning the competition between the onsite interaction $U$ and the superlattice potential depth $\Delta$, we identified three distinct quantum phases: a SF phase, a uniform Mott insulator with one boson per site (MI-I), and a sublattice-imbalanced Mott insulator (MI-II) featuring double occupation on one sublattice and vacancy on the other. We demonstrated that the SF–MI transitions are continuous and consistent with second-order quantum phase transitions. Moreover, we examined the thermal behavior of the SF phase, revealing the BKT transition from SF to normal fluid.
	
	Our results highlight the rich many-body physics enabled by superlattice-induced symmetry breaking in honeycomb geometries and offer theoretical guidance for future experiments aiming to realize and probe such phases in ultracold atom systems. Given the increasing experimental capability to engineer tunable optical superlattices, the honeycomb superlattice presents a realistic and promising platform for exploring correlated quantum phases beyond traditional lattice setups.

	\begin{acknowledgments}
		We acknowledge support from the National Natural Science Foundation of China (NSFC) under Grants No 12204173 and 12275002, and the University Annual Scientific Research Plan of Anhui Province under Grant No. 2022AH010013. 
	\end{acknowledgments}
	
	\bibliography{superlattice}
	
\end{document}